\documentclass[copyright,creativecommons]{eptcs}
\providecommand{\event}{PLACES 2026} % Name of the event you are submitting to

\usepackage{amsmath,amssymb,mathtools,amsthm}
\usepackage{stmaryrd}
\usepackage{mathpartir} % nice inference rules
\usepackage{xspace}
\usepackage{bm}
\usepackage{thmtools}
\usepackage{booktabs}
\usepackage{array}
\usepackage{hyperref}
\usepackage{prftree}
\usepackage{xcolor}
\usepackage{xurl}
\usepackage{doi}
\usepackage[numbers]{natbib}

\DeclareMathAlphabet{\mathscr}{OMS}{rsfs}{m}{n}

\newcommand{\titlerunning}{Towards MPST for Highly-Concurrent and Fault-Tolerant Web Applications}
\newcommand{\authorrunning}{Richard Casetta et al.}

\begin{document}

\title{Towards Multiparty Session Types for Highly-Concurrent and
Fault-Tolerant Web Applications}

\author{
Richard Casetta
\institute{BNP Paribas}
\institute{Univ. Grenoble Alpes, Inria, CNRS, Grenoble INP, LIG}
\email{}
Nils Gesbert
\institute{Univ. Grenoble Alpes, Inria, CNRS, Grenoble INP, LIG}
\email{}
Pierre Genevès
\institute{Univ. Grenoble Alpes, Inria, CNRS, Grenoble INP, LIG}
\email{}
}

\maketitle

\begin{abstract}
Modern web applications combine persistent state updates, concurrent
interactions, and unreliable communication with external services.
Failures such as timeouts can occur after partial state changes,
producing temporary inconsistencies whose resolution depends on liveness
properties that are often not verified in practice. Although formal
methods offer rigorous guarantees for reasoning about complex software,
they remain rarely adopted in enterprise settings due to their perceived
complexity and lack of practical automation. Multiparty Session Types
(MPST) offer strong guarantees for communication safety, yet they do not
account for the interplay between state evolution, dynamic workflow
structure, and failure behaviour that are essential for reasoning about
the correctness of real web applications. This paper introduces a
global‑type framework that equips MPST with explicit failure semantics
and dynamic participation. We define the syntax and operational
semantics of these enriched global types and establish core properties,
including coherence preservation. This foundation enables formal
reasoning about communications in web applications where failures may
occur, and lays the groundwork for future stateful extensions and
automated verification of liveness properties.
\end{abstract}

\newcommand{\tto}[1]{\xrightarrow{\;\!\!#1\!\!\;}}
\newcommand{\cto}{\mathrel{\mkern1mu\to\mkern1mu}}
\newcommand{\ncto}{\mathrel{\mkern1mu\not\to\mkern1mu}}
\newcommand{\kw}[1]{\textsf{#1}}
\newcommand{\nt}[1]{\mathit{#1}}
\newcommand{\id}[1]{\mathtt{#1}}
\newcommand{\set}[1]{\{\,#1\,\}}
\newcommand{\angles}[1]{\langle #1 \rangle}
\newcommand{\hl}[1]{\colorbox{yellow}{$\displaystyle #1$}}
\newcommand{\thl}[1]{\colorbox{yellow}{#1}}

\newcommand{\Role}{\mathcal{T}}
\newcommand{\ch}[1]{\mathsf{#1}}
\newcommand{\lab}[1]{\mathsf{#1}}
\newcommand{\ptype}[1]{\mathsf{#1}}
\newcommand{\ep}{\mathsf{ep}}
\newcommand{\End}{\kw{end}}
\newcommand{\gstop}{\kw{stop}}
\newcommand{\X}{\kw{X}}

\newcommand{\role}[1]{\mathsf{#1}}
\newcommand{\thread}[2]{\role{#1}\!\left[#2\right]}

\newcommand{\tout}{\kw{timeout}}
\newcommand{\trans}[3]{#1 \tto{\;#2\;} #3}
\newcommand{\G}{G}
\newcommand{\act}{\gamma}
\newcommand{\gdot}{\,.\,}
\newcommand{\gchoice}[2]{\sum\limits_{#1} #2}
\newcommand{\gpar}[2]{#1 \mid\mid #2}
\newcommand{\gmu}[2]{\mu\,#1\,.\,#2}
\newcommand{\gvar}[1]{#1}
\newcommand{\gend}{\End}
\newcommand{\bor}{\;\Bigm|\;}

\newcommand{\gsend}[5]{#1 \cto #2 : \ch{#3} \{\, \lab{#4}(\ptype{#5}) \,\}}
\newcommand{\gtimeout}[3]{#1 \cto #2 : \kw{timeout}\; \ch{#3}}
\newcommand{\gnew}[3]{#1 \cto #2 : \ch{\eta}(\nu\, \ch{#3})}
\newcommand{\gconnerr}[1]{#1 : \kw{connErr}}
\newcommand{\gstate}[2]{#1\langle\, #2 \,\rangle}

\newcommand{\gsendb}[4]{#1 \cto #2 : \ch{#3} \langle #4 \rangle}
\newcommand{\gsenderr}[4]{#1 \ncto #2 : \ch{#3} \langle #4 \rangle}
\newcommand{\mpayload}[2]{#1(#2)}
\newcommand{\mtimeout}{\kw{timeout}}
\newcommand{\mnew}[1]{\kw{new}(#1)}
\newcommand{\mconnerr}{\kw{connErr}}
\newcommand{\newp}{\nu_p}
\newcommand{\cL}{\mathcal{L}}

\newcommand{\T}{\role{t}}
\newcommand{\tp}{\role{p}}
\newcommand{\utp}{\role{p}^\dagger}
\newcommand{\ctp}{\role{p}^\lightning}
\newcommand{\tq}{\role{q}}
\newcommand{\utq}{\role{q}^\dagger}
\newcommand{\ctq}{\role{q}^\lightning}
\newcommand{\uT}{\role{t}^\dagger}
\newcommand{\cT}{\role{t}^\lightning}
\newcommand{\Tp}{\role{t'}}
\newcommand{\U}{\ptype{U}}
\newcommand{\Schan}{\ch{s}}
\newcommand{\Tchan}{\ch{t}}
\newcommand{\Rchan}{\ch{r}}
\newcommand{\labl}{\lab{l}}

\newcommand{\gact}[2]{#1 \set{#2}}
\newcommand{\gcomm}[3]{#1 \tto{} #2: #3}
\newcommand{\e}{\mathcal{E}}

\newcommand{\subj}{\mathsf{subj}}
\newcommand{\init}{\mathsf{init}}
\newcommand{\pt}{\mathsf{pt}}
\newcommand{\chan}{\mathsf{chan}}

\newcommand{\actnew}[3]{#1\,\nu\,#2: \ch{#3}}
\newcommand{\actsend}[4]{#1 \cto #2 : #3 \set{#4}}
\newcommand{\actsto}[3]{#1#2!#3}
\newcommand{\actrto}[3]{#1#2?#3}
\newcommand{\acterr}[2]{#1 \not\tto{} #2: \e}
\setlength\fboxsep{0pt}

\newcommand{\server}{\kw{server}}
\newcommand{\client}{\kw{client}}
\newcommand{\user}{\kw{user}}
\newcommand{\api}{\kw{api}}

\theoremstyle{definition}
\newtheorem{definition}{Definition}
\newtheorem{example}{Example}[section]

\theoremstyle{theorem}
\newtheorem{theorem}{Theorem}
\newtheorem{lemma}[theorem]{Lemma}

\section{Introduction}\label{introduction}

Modern web applications orchestrate workflows that interleave
user-interface actions, backend logic, persistent state updates, and
calls to unreliable external services. Failures such as timeouts or
connection errors frequently arise after partial state changes, leading
to temporary inconsistencies that standard Multiparty Session Types
(MPST) cannot express.

\subsection{\texorpdfstring{Motivating example
\label{sec:motivating-example}}{Motivating example }}\label{motivating-example}

Consider a user purchasing a ticket for her favourite show through a
typical web application. Such applications integrate several layers: a
client handling user interactions, a backend server orchestrating
multiple concurrent requests, and external services (e.g., payment,
inventory, fraud‑prevention APIs) operated by independent teams.

A simplified ticket‑ordering workflow is as follows: the user submits an
order through the web client, the client forwards the request to the
backend, then continues handling further user interactions, the backend
records the order and forwards payment and inventory requests to
external services and once these services respond, the backend returns a
final confirmation to the client, updating the displayed state.

Although the happy path appears straightforward, failures introduce
subtle inconsistencies. In practice, external APIs frequently suffer
from high load or unpredictable latency. When the payment or inventory
service responds too slowly, the backend may commit state locally (e.g.,
create the order, reserve seats) yet fail to reply to the client before
its timeout window elapses. The user then sees the familiar error
message: ``Something went wrong. Please try again later''. At this
point, two distinct states coexist: the backend state, where the order
may already be created and the seats reserved and the client state,
where the order is believed to have failed.

A simple page refresh often reconciles this discrepancy: the client
issues a fresh request and re-synchronizes with the authoritative
backend state, revealing whether the order was successfully recorded or
rejected (e.g., due to invalid payment or insufficient inventory).
Crucially, this behaviour expresses a liveness property: the user must
eventually obtain a coherent view of the order's outcome, even when
failures and state updates interleave. However, this property is not
guaranteed by the standard communication protocol alone. The resolution
relies on the existence of a reconciliation path (here, the manual
refresh) which bridges the inconsistent states. In analogous
server‑to‑server workflows, no such manual recovery step exists, and the
system may remain indefinitely inconsistent if the protocol does not
explicitly model and enforce a liveness‑preserving recovery path. This
simple yet ubiquitous example illustrates a core challenge of modern web
applications: state updates and failures routinely coexist, and
correctness depends on eventually reconciling their effects.

\subsection{Contributions}\label{contributions}

We introduce a global‑type framework extending MPST with explicit
timeout and connection‑error actions, dynamic thread spawning, and
crashes of active participants. We define the syntax and
labelled-transition semantics of these global types, together with
coherence conditions ensuring determinism of choices, timeout
completeness, and robustness to faults. We furthermore prove two
fundamental properties: preservation of coherence and absence of orphan
participants, ensuring that reductions maintain correctness and that all
actions involve valid, active endpoints. This framework provides a
foundation for reasoning about communication behaviours in web
applications and prepares the way for future work defining state updates
extension, local types, projection, and automated liveness verification.

In the following, Section\,2 develops the syntax and semantics;
Section\,3 presents key properties of the global type; Section 4 reviews
trade-offs and design choices; Section\,5 situates our work; Section\,6
concludes and mentions future work.

\section{Syntax and Semantics}\label{syntax-and-semantics}

We consider that participants are threads, represented by indexed roles
(\(A_{\mathsf{0}}, B_{\mathsf{1}}, ...\)). As such, we define
\(role(A_{\mathsf{r}}) = A\). As we use names \(\mathsf{p}, \mathsf{q}\)
for participants, if we have \(\mathsf{p}= A_{\mathsf{r}}\) we write
\(role(\mathsf{p})[\mathsf{s}]\) to mean \(A_\mathsf{s}\).

\subsection{Global type syntax}\label{global-type-syntax}

The syntax of global types is inspired from \citep{structured_com} with
annotated thread directly in the syntax and local thread spawning, and
we extend the crashed participants from \citep{barwell_generalised_2022}
to also model two-step timeouts on a channel. Let \(\mathcal{N}\) be a
set of channel names.

\[
\begin{tabular}{c c l l}
$G$    & $::=$   & $\mathsf{p}^{\colorbox{yellow}{$\displaystyle \dagger$}} \xrightarrow{\;\!\!\!\!\;} \mathsf{q}^{\colorbox{yellow}{$\displaystyle \dagger$}}: \mathsf{s} \{\,m_i \,.\,G_i\,\}_{i\in I}$ & Action \\
        & $\;\Bigm|\;$  & $\sum\limits_{i} G_i$                                                   & Choice\\
        & $\;\Bigm|\;$  & $G_1 \mid\mid G_2$                                                   & Parallel composition \\
        & $\;\Bigm|\;$  & $\mu\,\textsf{X}\,.\,G \;\Bigm|\;\textsf{X}\;\Bigm|\;\textsf{end}$                     & Recursion, type variable, termination
\end{tabular}
\]

\begin{mathparpagebreakable}
\mathsf{s} \in \mathcal{N} \cup \{\,\tau\,\}

m = l(T) \;\Bigm|\;\nu \mathsf{s}\;\Bigm|\;\mathcal{E}

\colorbox{yellow}{$\displaystyle \dagger$} = \cdot \;\Bigm|\;\colorbox{yellow}{$\displaystyle \lightning$}
\end{mathparpagebreakable}

In this model, actions
\(\mathsf{p}^{\colorbox{yellow}{$\displaystyle \dagger$}} \xrightarrow{\;\!\!\!\!\;} \mathsf{q}^{\colorbox{yellow}{$\displaystyle \dagger$}}: \mathsf{s} \{\,m_i \,.\,G_i\,\}\)
represents either a message communication \(m = l(T)\) over the private
channel \(\mathsf{s}\in \mathcal{N}\), a connection request
\(m = \nu \mathsf{t}\) when \(\mathsf{p}\neq \mathsf{q}\) over the
public channel \(\mathsf{s}\), a local thread‑spawning event
\(m = \nu \mathsf{t}\) when \(\mathsf{p}= \mathsf{q}\) over \(\tau\)
channel or an error handling branch \(m = \mathcal{E}\). Concretely, an
\(\mathcal{E}\) branch for a communication represents a timeout, while
it represents a connection error when it is a request.
\(\colorbox{yellow}{$\displaystyle \dagger$}\) is a runtime annotation
to mark whether endpoints are still alive (\(\cdot\)) or not
(\(\lightning\)). These error actions express that a communication could
not proceed, ensuring that failures are explicitly represented in the
global specification. The initiator of an action is the first
participant
(\(\mathsf{init}(\mathsf{p}^\dagger \xrightarrow{\;\!\!\!\!\;} \mathsf{q}^\dagger: \mathsf{s} \{\,m_i \,.\,G_i\,\}) = \mathsf{p}\)).
For choices, we define \(\mathsf{init}\) only when all branches have the
same initiator. We also note \(pub(G)\) the fixed set of participants
that expose a public channel. These participants cannot appear elsewhere
than in the receiver of a request. Finally, recursion is guarded and
highlighted parts of the syntax defines runtime types. We define a
congruence relation (\(\equiv\)) on global types by:

\begin{mathpar}
G_1 \mid\mid G_2 \equiv G_2 \mid\mid G_1

G \mid\mid \textsf{end} \equiv G
\end{mathpar}

\begin{example}
We illustrate the benefit of our framework using a simple purchase workflow.
A classical MPST specification for this interaction is:

$$
\begin{aligned}
G_{purchase}^{classic} = &\;\textsf{server} \xrightarrow{\;\!\!\!\!\;} \textsf{api}: \mathsf{s} \{\,\textsf{Purchase}(Order)\,\} \,.\,\\
&\;\textsf{api} \xrightarrow{\;\!\!\!\!\;} \textsf{server}: \mathsf{s} \{\,\textsf{OrderPurchased}(Id)\,\} \,.\,\\
&\; \textsf{server} \xrightarrow{\;\!\!\!\!\;} \textsf{client}: \mathsf{t} \{\,\textsf{OrderComplete}(Infos)\,\} \,.\,\textsf{end}
\end{aligned}
$$

This well-formed global type describes the happy path only: it provides no way to express that the API may time out or become temporarily unavailable.
Our framework introduces \emph{explicit timeout actions} that make such behaviours part of the protocol:

$$
\begin{aligned}
G_{purchase} =\, &\textsf{server} \xrightarrow{\;\!\!\!\!\;} \textsf{api}: \mathsf{s} \{\,\textsf{Purchase}(Order)\,\} \,.\,\\
&\textsf{api} \xrightarrow{\;\!\!\!\!\;} \textsf{server}: \mathsf{s} \left\{\begin{aligned}
& \textsf{OrderPurchased}(Id) \,.\,\textsf{server} \xrightarrow{\;\!\!\!\!\;} \textsf{client}: \mathsf{t} \{\,\textsf{OrderComplete}(Infos)\,\} \,.\,\textsf{end}\\
& \mathcal{E}\,.\,\textsf{server} \xrightarrow{\;\!\!\!\!\;} \textsf{client}: \mathsf{t} \{\,\textsf{UnexpectedError}\,\} \,.\,\textsf{end}^{*} \\
\end{aligned}
\right.
\end{aligned}
$$

The second branch models an \emph{asynchronous timeout}: it may occur even if the API has not crashed, capturing the desynchronisation often observed in web applications.
This goes beyond crash-stop behaviour and allows us to express failures that arise from latency, user impatience, or premature cancellation.

Finally, our framework can combine timeouts, crashes, and dynamic participants, enabling a protocol to \emph{restart} after a failure.
For instance, by replacing the trailing $\textsf{end}^{*}$ in the timeout branch with a recursion variable~$\textsf{X}$, we obtain:

$$
\begin{aligned}
G_{restart} = & \mu\,\textsf{X}\,.\,\textsf{client} \xrightarrow{\;\!\!\!\!\;} \textsf{server}: \mathsf{k}_{server} \{\,\nu \mathsf{t}\,\} \,.\,G_{purchase}\{ \textsf{end}^{*} / X\}
\end{aligned}
$$

A new server thread is spawned on each iteration on the public endpoint $\mathsf{k_{server}}$; if it times out or crashes, the client may simply re-enter the loop.
This models the familiar \emph{refresh} behaviour of web clients.
\end{example}

\begin{definition}[Live participant set]
The live participant set of an interaction over a set of participant S, written $\mathcal{L}(S)$ is defined by:
\begin{mathparpagebreakable}
\mathcal{L}(S) = \{\,\mathsf{p}^\dagger \in S \mid \dagger = \cdot\,\}
\end{mathparpagebreakable}
\end{definition}

\begin{definition}[Participants]
The \emph{participants} of a global type $G$ are the indexed roles that appear as live endpoints of its actions, representing exactly the threads that actually take part in the protocol.
\begin{mathparpagebreakable}
\footnotesize
pt(\textsf{X}) = pt(\textsf{end}) = \emptyset

pt(\mu\,\textsf{X}\,.\,G) = pt(G)

pt(G_1 \mid\mid G_2) = pt(G_1) \cup pt(G_2)

pt(\sum\limits_{i \in I} G_i) = \bigcup\limits_{i \in I} pt(G_i)

pt(\mathsf{p}^\dagger \xrightarrow{\;\!\!\!\!\;} \mathsf{q}^\dagger: \mathsf{s} \{\,m_i \,.\,G_i\,\}_{i\in I_\mathcal{E}}) = \bigcup\limits_{i \in I} \sigma(\mathsf{p}^\dagger \xrightarrow{\;\!\!\!\!\;} \mathsf{q}^\dagger: \mathsf{s} \{\,m_i \,.\,G_i\,\})

\sigma(\mathsf{p}^\dagger \xrightarrow{\;\!\!\!\!\;} \mathsf{q}^\dagger: \mathsf{s} \{\,m \,.\,G\,\}) = \mathcal{L}(\{\,\mathsf{p}^\dagger, \mathsf{q}^\dagger\,\}) \cup (pt(G) \setminus \nu_p(\mathsf{q}, m))
\end{mathparpagebreakable}
where $\nu_p(\mathsf{q}, \nu \mathsf{t}) = \{\,role(\mathsf{q})[\mathsf{t}]\,\}$ and $\nu_p(\mathsf{q}, m) = \emptyset$ otherwise represents the new participant set.
\end{definition}

Intuitively, a thread waiting to be spawned should not yet be considered
crashable; hence we remove the newly created role instance from the
active set. This allows us to characterise precisely which participants
may fail at a given point in the protocol. We can now define the crash
of a participants.

\begin{definition}[Crash] We define the function $G\lightning\mathsf{r}$ that crashes $\mathsf{r}$ over the global type $G$:

\begin{mathparpagebreakable}
\footnotesize
\setlength{\jot}{-.2cm}
\textsf{end}\lightning\mathsf{r} = \textsf{end}

\textsf{X}\lightning\mathsf{r} = \textsf{X}

(\mu\,\textsf{X}\,.\,G)\lightning\mathsf{r} = \mu\,\textsf{X}\,.\,(G\lightning\mathsf{r})

\left(\sum\limits_{i \in I} G_i\right)\lightning\mathsf{r} = \sum\limits_{i \in I} \left(G_i\lightning\mathsf{r}\right)

(G_1 \mid\mid G_2)\lightning\mathsf{r} = (G_1\lightning\mathsf{r}) \mid\mid (G_2\lightning\mathsf{r})

(\mathsf{p}^\dagger \xrightarrow{\;\!\!\!\!\;} \mathsf{q}^\dagger: \mathsf{s} \{\,m_i \,.\,G_i\,\}_{i\in I_\mathcal{E}})\lightning\mathsf{r} = \begin{cases}
  \mathsf{p}^\dagger \xrightarrow{\;\!\!\!\!\;} \mathsf{q}^\dagger: \mathsf{s} \{\,m_i \,.\,(G_i\lightning\mathsf{r})\,\}_{i\in I_\mathcal{E}} & \mathsf{r} \not\in \{\,\mathsf{p}, \mathsf{q}\,\} \\
  \mathsf{p}^\lightning \xrightarrow{\;\!\!\!\!\;} \mathsf{q}^\dagger: \mathsf{s} \{\,\mathcal{E}\,.\,(G_\mathcal{E}\lightning\mathsf{r})\,\} & \mathsf{p}= \mathsf{r} \neq \mathsf{q}\\
  \mathsf{p}^\dagger \xrightarrow{\;\!\!\!\!\;} \mathsf{q}^\lightning: \mathsf{s} \{\,\mathcal{E}\,.\,(G_\mathcal{E}\lightning\mathsf{r})\,\} & \mathsf{p}\neq \mathsf{r} = \mathsf{q}\\
  \mathsf{p}^\lightning \xrightarrow{\;\!\!\!\!\;} \mathsf{q}^\lightning: \mathsf{s} \{\,\mathcal{E}\,.\,(G_\mathcal{E}\lightning\mathsf{r})\,\} & \mathsf{p}= \mathsf{r} = \mathsf{q}\\
\end{cases}
\end{mathparpagebreakable}
\end{definition}

\begin{example}
To illustrate the semantics of crash propagation in global types, we define the global type derived from $G_{purchase}$ after the $\textsf{server}$ send:
$$
\small
G_{purchase}^{\textsf{response}} = \textsf{api} \xrightarrow{\;\!\!\!\!\;} \textsf{server}: \mathsf{s} \begin{cases}
\textsf{OrderPurchased}(Id) \,.\,\textsf{server} \xrightarrow{\;\!\!\!\!\;} \textsf{client}: \mathsf{t} \{\,\textsf{OrderComplete}(Infos)\,\} \,.\,\textsf{end}\\
\mathcal{E}\,.\,\textsf{server} \xrightarrow{\;\!\!\!\!\;} \textsf{client}: \mathsf{t} \{\,\textsf{UnexpectedError}\,\} \,.\,\textsf{end}\\
\end{cases}
$$
In this case, we have:
$$
G_{purchase}^{\textsf{response}}\lightning\textsf{api}= \textsf{api}^\lightning \xrightarrow{\;\!\!\!\!\;} \textsf{server}: \mathsf{s} \{\,\mathcal{E}\,.\,\textsf{server} \xrightarrow{\;\!\!\!\!\;} \textsf{client}: \mathsf{t} \{\,\textsf{UnexpectedError} \,.\,\textsf{end}\,\}\,\}
$$
The crash of the $\textsf{api}$ removes all non-error branches and marks which participant has failed.
\end{example}

\subsection{Global type semantics}\label{global-type-semantics}

We give a Labelled Transition System (LTS) semantics to global types
with the following rules, where actions are defined by:

\begingroup
\footnotesize
\begin{center}
\begin{tabular}{rclclcl}
$\alpha ::=$ & $\mathsf{p} \mathrel{\mkern 1mu\to\mkern 1mu}\mathsf{q} : \mathsf{s} \{\,m\,\}$ where $(m\neq\mathcal{E})$ & $\;\Bigm|\;$ & $\mathsf{p} \not\xrightarrow{\;\!\!\!\!\;} \mathsf{s}: \mathcal{E}$ & $\;\Bigm|\;$ & $\mathsf{p}\lightning$ \\[4pt]
&$\mathsf{subj}(\cdot)=\{\,\mathsf{p},\mathsf{q}\,\}$ &&$\mathsf{subj}(\cdot)=\{\,\mathsf{p}\,\}$ &&$\mathsf{subj}(\cdot)=\{\,\mathsf{p}\,\}$
\end{tabular}
\end{center}
\endgroup

\begin{definition}[LTS semantics]
The LTS semantics of global types is defined by:

\begin{mathpar}
\footnotesize
\prftree[l]
{[\textsc{com}]}
{j \in I}
{m_j \neq \mathcal{E}}
{\mathsf{p} \xrightarrow{\;\!\!\!\!\;} \mathsf{q}: \mathsf{s} \{\,m_i \,.\,G_i\,\}_{i\in I_\mathcal{E}} \xrightarrow{\;\!\!\mathsf{p} \mathrel{\mkern 1mu\to\mkern 1mu}\mathsf{q} : \mathsf{s} \{\,m_j\,\}\!\!\;} G_j}

\prftree[l]
{[\textsc{com$\lightning$!}]}
{\mathsf{p}\neq \mathsf{q}}
{\mathsf{p} \xrightarrow{\;\!\!\!\!\;} \mathsf{q}^\dagger: \mathsf{s} \{\,m_i \,.\,G_i\,\}_{i\in I_\mathcal{E}} \xrightarrow{\;\!\!\mathsf{p} \not\xrightarrow{\;\!\!\!\!\;} \mathsf{s}: \mathcal{E}\!\!\;} \mathsf{p}^\lightning \xrightarrow{\;\!\!\!\!\;} \mathsf{q}^\dagger: \mathsf{s} \{\,\mathcal{E}\,.\,G_\mathcal{E}\,\}}

[\textsc{com$\lightning$?}]\;
{\mathsf{p}^\dagger \xrightarrow{\;\!\!\!\!\;} \mathsf{q}: \mathsf{s} \{\,l_i(T_i) \,.\,G_i, \mathcal{E}\,.\,G_\mathcal{E}\,\}_{i\in I} \xrightarrow{\;\!\!\mathsf{q} \not\xrightarrow{\;\!\!\!\!\;} \mathsf{s}: \mathcal{E}\!\!\;} \mathsf{p}^\dagger \xrightarrow{\;\!\!\!\!\;} \mathsf{q}^\lightning: \mathsf{s} \{\,\mathcal{E}\,.\,G_\mathcal{E}\,\}}

\prftree[l]{[\textsc{concur}]}
{\forall i\in I_\mathcal{E}}
{G_i \xrightarrow{\;\!\!\alpha\!\!\;} G_i' \quad\mathsf{subj}(\alpha) \cap \left(\mathcal{L}(\{\,\mathsf{p}^\dagger, \mathsf{q}^\dagger\,\}) \cup \nu_p(\mathsf{q}, m_i)\right) = \emptyset}
{\mathsf{p}^\dagger \xrightarrow{\;\!\!\!\!\;} \mathsf{q}^\dagger: \mathsf{s} \{\,m_i \,.\,G_i\,\}_{i\in I_\mathcal{E}} \xrightarrow{\;\!\!\alpha\!\!\;} \mathsf{p}^\dagger \xrightarrow{\;\!\!\!\!\;} \mathsf{q}^\dagger: \mathsf{s} \{\,m_i \,.\,G_i'\,\}_{i\in I_\mathcal{E}}}

\prftree[l]{[\textsc{$\lightning$}]}
{\mathsf{p}\in \mathsf{pt}(G)}
{G\xrightarrow{\;\!\!\mathsf{p}\lightning\!\!\;} G\lightning\mathsf{p}}

\prftree[l]{[\textsc{par}]}
{G_1 \xrightarrow{\;\!\!\alpha\!\!\;} G'_1}
{G_1 \mid\mid G_2 \xrightarrow{\;\!\!\alpha\!\!\;} G'_1 \mid\mid G_2}

\prftree[l]{[\textsc{rec}]}
{G\{ \mu\,\textsf{X}\,.\,G / \textsf{X}\} \xrightarrow{\;\!\!\alpha\!\!\;} G'}
{\mu\,\textsf{X}\,.\,G \xrightarrow{\;\!\!\alpha\!\!\;} G'}

\prftree[l]{[\textsc{choice}]}
{j\in I}
{G_j \xrightarrow{\;\!\!\alpha\!\!\;} G'_j \quad init\!\left(\sum\limits_{i \in I} G_i\right) \in\mathsf{subj}(\alpha)}
{\sum\limits_{i \in I} G_i \xrightarrow{\;\!\!\alpha\!\!\;} G'_j}

\prftree[l]{[\textsc{choice-br}]}
{\forall i \in I}
{G_i \xrightarrow{\;\!\!\alpha\!\!\;} G'_i}
{\sum\limits_{i \in I} G_i \xrightarrow{\;\!\!\alpha\!\!\;} \sum\limits_{i \in I} G'_i}
\end{mathpar}

Rule [\textsc{com}] models synchronous communication and spawning (either a request or locally).
Rule [\textsc{com$\lightning$!}] allows the sender to handle a timeout or a connection error.
Rules [\textsc{com$\lightning$?}] allows the receiver of a message to stop waiting and timeout before communication takes place, only if it is not a server ($\mathsf{p}\neq \mathsf{q}$).
Rule [\textsc{$\lightning$}] represents the crash of an active participant.
Rule [\textsc{choice}] models the resolution of the choice by its initiator.
Rule [\textsc{par}] allows the first global type to reduce, which models concurrent nature of the parallel composition as the second global type can also reduce due to congruence.
Rule [\textsc{rec}] represents the unfolding.
Rule [\textsc{concur}] ensures that unrelated actions can be performed before an action as long as it does not interfere with the remaininging live participants.
Rule [\textsc{choice-br}] allows an action to be made before a choice is resolved if it is available in all branches of a choice.

While the core semantics is synchronous, non-blocking sends are obtained by spawning local threads handling the communication, much like goroutines in Go, allowing the parent thread to continue and synchronize later.
Finally, we define [\textsc{rec}] up to $\alpha$-equivalence, in order to create new bound channel names when unfolding to avoid name clash.
\end{definition}

\begin{example}
Now that the semantics rules are defined, we can show how the asynchronous timeout works.
Starting from $G_{purchase}^{\textsf{response}}$, we have, at this point, several possible actions:
(i) the $\textsf{api}$ may resolve the choice by selecting one of the branches;
(ii) the $\textsf{server}$ or the $\textsf{client}$ may time out before the choice is committed;
(iii) one of the active participants may crash.

In the case where the $\textsf{server}$ times out, we have:
$$
\small
G_{purchase}^{\textsf{response}} \xrightarrow{\;\!\!\textsf{server} \not\xrightarrow{\;\!\!\!\!\;} \mathsf{s}: \mathcal{E}\!\!\;} \textsf{api} \xrightarrow{\;\!\!\!\!\;} \textsf{server}^\lightning: \mathsf{s} \begin{cases}
\mathcal{E}\,.\,\textsf{server} \xrightarrow{\;\!\!\!\!\;} \textsf{client}: \mathsf{t} \{\,\textsf{UnexpectedError}\,\} \,.\,\textsf{end}\}\\
\end{cases}
$$
Together, these rules provide an asynchronous timeout for active participants.
Note also that crashing a participant is operationally equivalent to performing a single action that triggers an error on every channel in which that participant is involved (private or public).
\end{example}

Let \(\Delta; \Gamma\) be environments where
\(\Delta = \emptyset \mid \Delta, \textsf{X}: \Gamma\) is used for
recursive variable tracking existing channels, and
\(\Gamma = \emptyset \mid \Gamma, \mathsf{s}: \mathsf{p}\mid \Gamma, \mathsf{s}: \{\,\mathsf{p}, \mathsf{q}\,\}\)
is used for public and private channels.

\begin{definition}[Coherence]
We say that a global type is coherent it its channel usage and participants index are coherent.
Concretely, a global type $G$ is coherent if there exists $\Delta, \Gamma$ such that $\Delta; \Gamma \vdash G$ where the rule of the judgement are given by:

\begin{mathparpagebreakable}
\footnotesize
\prftree[l]
{[\textsc{send}]}
{\Delta;\Gamma \vdash G_\mathcal{E}}
{\forall i\in I,\; \Delta;\Gamma, \mathsf{s}: \{\,\mathsf{p}, \mathsf{q}\,\} \vdash G_i}
{\forall i \neq j, l_i \neq l_j}
{\Delta; \Gamma, \mathsf{s}: \{\,\mathsf{p}, \mathsf{q}\,\} \vdash \mathsf{p} \xrightarrow{\;\!\!\!\!\;} \mathsf{q}: \mathsf{s} \{\,l_i(T_i) \,.\,G_i, \mathcal{E}\,.\,G_\mathcal{E}\,\}_{i\in I}}

\prftree[l]
{[\textsc{req}]}
{\mathsf{p}\neq \mathsf{q}}
{\mathsf{s}\neq \tau}
{\Gamma(\mathsf{s}) = \mathsf{q}}
{\mathsf{t} \notin \Gamma}
{\Delta;\Gamma, \mathsf{t}:\{\,\mathsf{p}, role(\mathsf{q})[\mathsf{t}]\,\} \vdash G}
{\Delta;\Gamma \vdash G_\mathcal{E}}
{\Delta;\Gamma \vdash \mathsf{p} \xrightarrow{\;\!\!\!\!\;} \mathsf{q}: \mathsf{s} \{\,\nu \mathsf{t} \,.\,G, \mathcal{E}\,.\,G_\mathcal{E}\,\}}

\prftree[l]
{[\textsc{spawn}]}
{\mathsf{t} \notin \Gamma}
{\Delta;\Gamma, \mathsf{t}:\{\,\mathsf{p}, role(\mathsf{p})[\mathsf{t}]\,\} \vdash G}
{\Delta;\Gamma \vdash G_\mathcal{E}}
{\Delta;\Gamma \vdash \mathsf{p} \xrightarrow{\;\!\!\!\!\;} \mathsf{p}: \tau \{\,\nu \mathsf{t} \,.\,G, \mathcal{E}\,.\,G_\mathcal{E}\,\}}

\prftree[l]
{[\textsc{fail}]}
{(\mathsf{q}\in \Gamma(\mathsf{s}) \lor \mathsf{p}= \mathsf{q}) \iff \Gamma' = \Gamma}
{(\mathsf{q}\not\in \Gamma(\mathsf{s}) \land \mathsf{p}\neq \mathsf{q}) \iff \Gamma' = \Gamma, s: \{\,\mathsf{p}^{\dagger_\mathsf{p}}, \mathsf{q}^{\dagger_\mathsf{q}}\,\}}
{\Delta;\Gamma \vdash G_\mathcal{E}}
{\Delta;\Gamma' \vdash \mathsf{p}^{\dagger_\mathsf{p}} \xrightarrow{\;\!\!\!\!\;} \mathsf{q}^{\dagger_\mathsf{q}}: \mathsf{s} \{\,\mathcal{E}\,.\,G_\mathcal{E}\,\}}

\prftree[l]
{[\textsc{par}]}
{\Delta_1;\Gamma_1 \vdash G_1}
{\Delta_2;\Gamma_2 \vdash G_2}
{\Delta_1 \cap \Delta_2 = \emptyset}
{\Gamma_1 \cap \Gamma_2 = pub(G_1) \cup pub(G_2)}
{\Delta_1, \Delta_2; \Gamma_1,\Gamma_2 \vdash G_1 \mid\mid G_2}

\prftree[l]
{[\textsc{sum}]}
{\forall i\in I,\; \Delta;\Gamma \vdash G_i}
{\exists \mathsf{p},\; \mathsf{init}(\sum\limits_{i \in I} G_i) = \mathsf{p}}
{\Delta;\Gamma \vdash \sum\limits_{i \in I} G_i}

\prftree[l]
{[\textsc{rec}]}
{\Delta, \textsf{X}: \Gamma;\Gamma \vdash G}
{\Delta; \Gamma \vdash \mu\,\textsf{X}\,.\,G}

\prftree[l]
{[\textsc{var-rec}]}
{\Delta, \textsf{X}: \Gamma; \Gamma \vdash \textsf{X}}

\prftree[l]
{[\textsc{end}]}
{\Delta;\emptyset \vdash \textsf{end}}
\end{mathparpagebreakable}
Rule [\textsc{send}] allows communication as long as the channel has been previously opened.
Rule [\textsc{req}] ensures the channel name is fresh and checks that it is a request to a public endpoint.
Rule [\textsc{spawn}] is the same but for local spawning.
Rule [\textsc{fail}] ensures that failure-handling branch are coherent and that the prefix has the channel state in its environment if it is a communication.
[\textsc{rec}] and [\textsc{var-rec}] stores the state of channels before the recursion and ensures that none of them has been deleted and all channels created during the recursion have been closed (timeout in our framework).
[\textsc{par}] ensures that only public channels are shared between environments of parallel global types, otherwise no channels, participants, recursion variables and private channels stored in recursion variables can be shared.
Rule [\textsc{end}] ensures that every channel has been closed.
[\textsc{sum}] allow a choice to be coherent if all branches are coherent.
We now show the properties satisfied by our coherence rules.
\end{definition}

\section{Global Type Properties}\label{global-type-properties}

We now state the key meta‑theoretic properties of our framework. All
proofs are given in \citep{Casetta2026WebMPSTExtended}. We begin by
establishing a basic structural property of the typing environment.

\begin{lemma}[Weakening] \label{lem:weak}
If $\Delta;\Gamma \vdash G$ and $\textsf{X}\notin \textsf{dom}(\Delta)$, then $\Delta, \textsf{X}:\Gamma';\Gamma \vdash G$
\end{lemma}

Weakening is essential in order to add variables binding in the
environment for a branch which does not use the recursion variable. We
then turn to ensuring that the semantics never enables actions from
invalid participants.

\begin{theorem}[No Orphan Participants] \label{thm:noop}
Let $G$ be a coherent global type.
Then for any $\alpha$ such that $G\xrightarrow{\;\!\!\alpha\!\!\;}$, $\mathsf{subj}(\alpha) \in pt(G)$.
\end{theorem}

This property ensures that every action in the protocol can always be
attributed to a currently active role, preventing actions from a crashed
or nonexistent role. We then show that crashes preserves coherence.

\begin{lemma}[Preservation under crash] \label{lem:wfc}
Let $G$ be a coherent global type.
Then for all $\mathsf{p}\in pt(G)$, $G\lightning\mathsf{p}$ is coherent.
\end{lemma}

This ensures that eliminating an active participant does not compromise
the structure of the protocol. Having established stability under
crashes, we now consider stability under operational steps.

\begin{theorem}[Preservation of Coherence] \label{thm:wfp}
Let $G$ be a coherent global type.
Then for any $G', \alpha$ such that $G\xrightarrow{\;\!\!\alpha\!\!\;} G'$, $G'$ is coherent.
\end{theorem}

Together, these results show that coherent global types remain stable
under crashes and reductions, and that every step of the semantics
involves only valid, active participants. This ensures that our
framework provides a coherent and robust foundation for reasoning about
failure‑prone communication behaviours.

\section{Discussion}\label{discussion}

We now discuss the rationale behind our semantic choices and explain why
the development of a new MPST theory was necessary.

\subsection{Directed Acyclic Graph
Topology}\label{directed-acyclic-graph-topology}

Classic MPST theories typically assume that all participants can
communicate with each other. In contrast, our setting mirrors the
topology observed in web applications. Due to the nature of TCP/IP
communication, a client must first establish a connection with a server
before interacting with it. Moreover, a participant can only communicate
with the participant that created it, or with participants it has itself
created. As a consequence, if two participants exist, then either one
created the other and direct communication is possible, or direct
communication is impossible. This structure naturally induces a directed
acyclic graph (DAG) of participants, which restricts expressivity.
However, this restriction significantly simplifies the theory.

Additionally, this topology implies that communication relies on a set
of binary channels that must remain dual along all execution paths:
there is no interaction with third parties that could alter the type of
a channel. As a result, deadlocks caused by cyclic dependencies between
three or more participants cannot occur. This aligns well with the
behaviour of web applications, where such deadlocks are rare in
practice.

\subsection{Sender-driven choice}\label{sender-driven-choice}

We adopt sums of global types with multiple receivers, also known as
sender-driven choice. While uncommon in MPST theories, this design
choice is mandatory in our context. Web applications are exposed to high
traffic as well as to maliciously crafted requests. In such
environments, every input must be validated to ensure correctness. This
requirement underlies the fail-fast principle: inputs are validated
early, and failures are handled immediately.

When validation fails, web applications typically send an error response
to the client rather than continuing the interaction. Therefore, our
theory must support the ability for a participant to choose between
sending an error or continuing the protocol, possibly by issuing a new
request to an external service.

\subsection{Crash design}\label{crash-design}

There are two main approaches to modelling crashes or timeouts of
participants in a fault-tolerant setting. The first approach preserves
all branches of the protocol while marking the crashed participant as no
longer alive (\(\lightning\)). This approach has little impact on
projection and allows the standard subtyping relation to relate global
and local semantics. However, it complicates the operational semantics:
a dedicated concurrency rule is required to enforce that, once a
participant has crashed or timed out, only the failure-handling branch
can reduce, while all other branches are blocked. This complexity is
exacerbated by the coherence judgement, which would need to ensure that
the error-handling branch is coherent even though its coherence
environment differs fundamentally from that of the stuck branches.

The alternative approach consists in pruning all branches except the
failure-handling branch. This yields a particularly clean semantics, as
only a single, simple concurrency rule is needed. The judgement rules
are also simplified, since only correct executions with an error branch
need to be distinguished from actual errors. However, this choice
complicates the relation between global and local types. When a receiver
crashes or times out, the corresponding branches are pruned from the
global type. Consequently, the projection of the sender contains only
the error-handling branch, whereas the local configuration still
contains all branches, effectively reversing the usual subtyping
relation.

We chose to prioritise semantic clarity, even at the cost of more
difficult proofs of operational equivalence (as long as proofs are
possible), rather than adopting a more complex semantics and coherence
judgement with possibly simpler proofs. To relate global and local
types, a possible future work is to define a subtyping relation based on
an erasure function that removes all branches involving closed channels
before applying the standard subtyping relation. While the impossible
branches are pruned, it does not mean there is a need for a global
synchronization mechanism: the global type only represents the
possibilities left, locally each participant will only be able to time
out on the relevant channel, without distinguishing between a crash and
a closure.

\subsection{Explicit failure-handling
branch}\label{explicit-failure-handling-branch}

Our theory requires the presence of an explicit error-handling branch.
In a top-down approach, where the protocol designer specifies the global
type, this increases the specification burden. This cost can be
mitigated by providing default error-handling branches that simply
disconnect the relevant channels.

A more ambitious alternative is to use the theory in a bottom-up manner
with global type synthesis: reconstructing the global type from a set of
local types. In this setting, the DAG topology may reduce the complexity
of the synthesis algorithm. This approach also better reflects how web
applications are developed in practice: external services are typically
implemented by different teams and composed later. Writing local types
is therefore more natural, particularly since explicit error handling is
meaningful at the level of individual services.

\subsection{Unbounded number of
participants}\label{unbounded-number-of-participants}

Due to dynamic participant creation and recursion, our theory supports
an unbounded number of participants. We abstract away from
implementation details (resource management, garbage collection) as
these are typically handled by the framework or the tool used in web
applications development.

\section{Related Work}\label{related-work}

MPST were originally introduced as a global discipline for specifying
message‑passing protocols \citep{honda_multiparty_2008}, building on
earlier dyadic interaction types \citep{goos_types_1993}. A rich body of
work has since extended MPST along several axes.

\textbf{Error Handling.} Affine sessions and exception‑tolerant
communication \citep{mostrous_affine_2018, exceptional}, optional blocks
\citep{optional}, event‑driven fault‑tolerant MPST
\citep{viering_multiparty_2021}, and crash‑stop semantics
\citep{barwell_generalised_2022, barwell_crash-stop_2025, fmpst} extend
MPST with failure behaviours. Tool‑oriented systems for Rust support
cancellation, affinity, and timed failure handling
\citep{lagaillardie_stay_2022, hou_fearless_2024}. However, these works
lack dynamic participation: once a participant fails, the protocol
cannot be safely restarted.

\textbf{Dynamic Behaviour.} MPST extensions have explored connection
actions and evolving communication structures
\citep{structured_com, huisman_explicit_2017}, dynamic interleaving of
sessions \citep{hutchison_global_2008, multi_session}, and nested or
modular protocol structure \citep{hutchison_nested_2012}. More recent
work supports replication \citep{brun_multiparty_2025} and unbounded
participant creation \citep{castro-perez_dynamically_2023}. Yet, dynamic
evolution in these systems is not integrated with error handling: they
cannot model recovery paths, restartable workflows, or liveness under
partial failures.

\textbf{Refinements.} Refinement‑typed MPST frameworks
\citep{zhou_statically_2020}, specification‑agnostic refinement systems
\citep{vassor_refinements_2024}, and design‑by‑contract choreographies
\citep{gheri_design-by-contract_2022} enhance MPST with logical
constraints on values and control flow. These techniques are orthogonal
to our contribution and represent natural future extensions to integrate
with failure‑aware dynamic protocols in order to verify liveness
properties.

\section{Conclusion and Future Work}\label{conclusion-and-future-work}

This work introduced a global‑type framework that extends Multiparty
Session Types with explicit timeout and connection‑error actions,
dynamic thread spawning, and a precise notion of participants. These
additions enable the specification of communication behaviours that
arise routinely in web applications, where concurrency, partial
progress, and failures interact in ways not captured by existing MPST
formalisms. We defined the syntax, labelled‑transition semantics, and
coherence conditions of the framework, and established core
meta‑theoretic guarantees, including preservation of coherence and the
absence of orphan participants. Several directions remain open. A
natural next step is the development of projection to local types and
corresponding endpoint type systems, enabling static verification of
implementations that incorporate failures and dynamic participants.
Further work includes integrating stateful reasoning to capture the
interplay between communication, persistent updates, and recovery, as
well as developing automated analyses for liveness and consistency
properties. Ultimately, we aim to support tool‑assisted verification and
type‑safe implementations of realistic, fault‑tolerant web protocols.

\bibliographystyle{eptcs}
\bibliography{./main.bbl}

\end{document}